\documentclass[10pt,twocolumn]{article}

\usepackage{amsfonts}
\usepackage{amssymb}
\usepackage{amsmath}
\usepackage{color}
\usepackage{graphicx}
\usepackage[pdftex]{hyperref}
\usepackage{authblk}

\usepackage[usenames,dvipsnames,svgnames,table]{xcolor}

\begin{document}

\title{Simflowny 2: An upgraded platform for scientific modeling and simulation}



\author[1]{A.~Arbona \thanks{Corresponding Author: aarbona@iac3.eu}}
\author[1]{B.~Mi\~{n}ano}
\author[1]{A.~Rigo} 
\author[1]{C.~Bona}
\author[1,2]{C.~Palenzuela}
\author[1]{A.~Artigues}
\author[1]{C.~Bona-Casas}
\author[1,2]{J.~Mass\'o}
\affil[1]{IAC$\,^3$, Universitat de les Illes Balears, Cra. de Valldemossa km 7.5, 07122, Palma, Spain}
\affil[2]{Departament de F\'isica, Universitat de les Illes Balears and Institut d'Estudis Espacials de Catalunya, Palma, Baleares E-07122, Spain}


%

\maketitle

\begin{abstract}
Simflowny is an open platform which automatically generates parallel code of scientific dynamical models 
for different simulation frameworks. Here we present
major upgrades on this software to support an extended set of families of models, in particular:
i) a new generic family for partial differential equations, which can include spatial derivatives
of any order,
ii) a new family for agent based models to study complex phenomena --either on a spatial domain or on a graph--. Additionally we introduce a flexible graphical user interface (GUI) to accommodate these and future families of equations.
This paper describes the new GUI architecture and summarizes the formal representation and implementation of these new families, providing several validation results. 
\end{abstract}

{\bf PROGRAM SUMMARY}
  
\begin{small}
\noindent
{\em Program Title: }Simflowny                                 \\
{\em Licensing provisions: } Apache License, 2.0                      \\
{\em Programming language: } Java, C++ and JavaScript   \\
{\em Journal Reference of previous version:} Comput. Phys. Comm. 184 (2013) 2321--2331  \\
{\em Does the new version supersede the previous version?:} Yes  \\
{\em Reasons for the new version:} Additional features \\
{\em Summary of revisions:}\\
Expanded support for Partial Differential Equations.\\
Support for Agent Based Models.\\
New Graphical User Interface.\\
{\em Computer: }\\
  Simflowny runs in any computer with Docker \cite{dockerintro}, installation details can be checked in the documentation of Simflowny \cite{simflownydoc}. It can also be compiled from scratch in any Linux system, provided the requirements are properly installed following documentation indications.\\
  The generated code runs on any Linux platform ranging from personal workstations to clusters and parallel supercomputers.  \\
{\em Nature of problem:}\\
Simflowny generates code for numerical simulation for a wide range of models \\
{\em Solution method:}\\
  Any discretization scheme based on either Finite Volume Methods, Finite Difference Methods, or meshless methods for Partial Differential Equations.\\
  Agent Based Model simulations execute their own algorithm as set in their models. \\
{\em Additional comments:}\\
  The software architecture is easily extensible for future additional model families and simulation frameworks.\\
Full documentation is available in the wiki home of the Simflowny project \cite{simflownydoc}. \\  

\end{small}

\section{Introduction}

We present a significantly upgraded  version 2 of Simflowny~\cite{Arbona20132321},
an  open platform for scientific dynamical models, composed by a Domain Specific Language (DSL)  and a web based Integrated Development Environment (IDE), which automatically generates efficient parallel code for simulation frameworks. 

Originally, Simflowny was built to address the fact that
efforts towards a formalization of mathematical equations in 
simulation, at a more abstract level than the mere program source code, are 
still rare. Usually, the process that goes from an idea reflected in a set 
of equations to the final code is pure art, in the sense that it lacks 
formalization and general tools to ease it.

In version 2 we have extended this formalization of mathematical equations to a more general family of scientific model paradigms (for example, Agent Base Models). Simflowny 
is aimed to create a community contributed platform where some elements of 
the simulation flow for Initial Value Problems are formalized. This formalization allows the reuse and exchange of such elements, even outside the platform itself. 

Simflowny has a simple yet ambitious goal: a complete split of the physics (by introducing the concept of models and their associated problems) from the numerical 
methods necessary for a simulation (the discretization schemes), and the automatic generation of the final simulation code (where the parallel features of the chosen framework will be optimally leveraged, as well as other provided capabilities).

The DSL is based on an XML Schema Definition (XSD) representation. The XSD schemas
prescribe the structure of the XML documents for models, problems, and discretization schemes. Where algorithms are to be included in either of these XML documents, this is done through a specific markup language developed for Simflowny, called SimML (Simulation Markup Language). SimML constitutes a
full-blown rule specification language (technically speaking,
the language is Turing-complete, meaning you can create any possible
algorithm with it). To insert mathematical expressions in the algorithms and documents, Simflowny prescribes MathML, the standard markup language  for representing mathematical expressions.
 
In version 2 the DSL has been expanded and 
currently supports the following scientific model paradigms, or \emph{families}:
\begin{itemize}
    \item Partial Differential Equations (PDE) written in balance-law form. This family was the only one supported in version 1 of Simflowny.    
    The PDEs are written as an 
    evolution system containing only first order derivatives both in time and space, which allow users to use numerical schemes based on Finite Volume Methods to deal with shocks and discontinuities. Some examples of balance
    law systems include the wave equation, Maxwell equations, Einstein equations,... This family also allows parabolic terms like the ones appearing, for instance, in the Navier-Stokes equations.
    
    \item Generic PDE, a new family of models that allows
    almost arbitrary forms of PDE evolution equations, including spatial derivatives of any order. The only restriction is that
    the system must be still first order in time. One could write in this form all the examples in balance-law form but directly as second order system (in space). Additionally, it also includes many other equations like the heat equation, the Navier-Stokes-Korteweg equation (third order in space), the Cahn-Hilliard equation (fourth order in space) and the Phase-Field-Crystal equation (sixth order in space).
    
    \item Agent-Based Models (ABM), another new family of models which simulate the evolution of a system of multiple interacting agents. These models are often built in order to study the emergence of complex phenomena. Agents might live either on a spatial domain (Spatial ABM) or on a graph (ABM on a Graph). Additionally,
    Spatial ABM contain two subfamilies: Cellular Automata, i.e. 
    agents statically bound to a cell in a mesh, and kinematic ABM, 
    where agents roam freely in a meshless domain. Some well-known examples are the Ising and Collective Motion (flocking) models for the Spatial ABM, and the Voter model and Cash and Goods \cite{Razakanirina2010} for ABM on a Graph. 
\end{itemize}

From the computer science point of view, 
Simflowny is built on the well-known three-tier architecture:
\begin{itemize}
\item A presentation tier, implemented as a web browser based 
graphical user interface (GUI).
\item A logic tier, based on an application server.
\item A data tier, combining native XML databases with bulk data storage.
\end{itemize}

In Simflowny 2, the original Graphical User Interface (GUI) of version 1, written ad hoc for the original PDE family, has been completely redone to make it flexible enough to automatically accommodate new families of equations, both of PDE nature or otherwise (such as ABM).

The current procedure to generate code is similar to 
the previous version, although new
features have been introduced to allow for more flexibility
with the new families. Although the details vary among
the families, the process to convert a mathematical model 
into a numerical code can be split in four stages (see fig. \ref{fig:workflow}):

\begin{figure*}
  \begin{center}
  \includegraphics[width=\columnwidth]{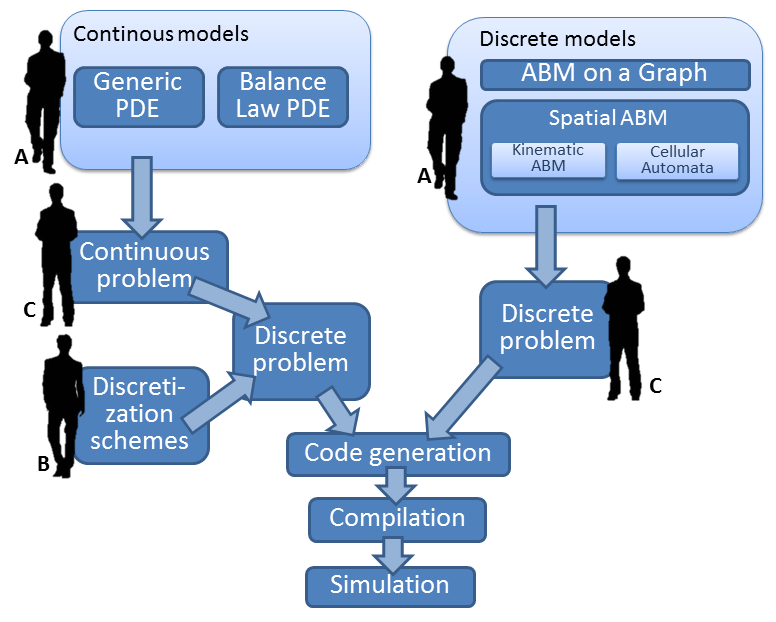}
  \caption{Workflow diagram, containing the following stages: 1) the mathematical {\bf model} containing either the evolution equations for PDEs, or a description of the interactions among the     multiple agents in ABM. 2) the {\bf problem}, which mainly includes the model, the domain of the simulation and the initial and boundary conditions. 3) 
  the {\bf discrete scheme}, which converts the continuous problem into a discrete one. Notice that ABM models are already discrete.
  4) the code generation, converting the discrete problem to code for the final framework. }
  \label{fig:workflow}
  \end{center}
\end{figure*}

\begin{itemize}
\item  The representation of the mathematical model, which  contains either the evolution equations to be evolved in the case of PDEs, or a description of the interactions among the multiple agents in ABM. 
\item The representation of the problem, which includes the mathematical model, the domain of the simulation, the analysis quantities and the initial and boundary conditions to be applied, either to the evolution fields (in PDEs) or to agent properties (in ABM). 
\item The representation of the discrete scheme, which converts the continuous problem into a discrete one by defining the space and time discretization operators
(for PDEs). This stage does not apply to the ABM families since the original models are already discrete.
\item The generation of the code from the discrete problem
to the final framework. These frameworks will essentially play the role of a mesh/memory manager by setting the domain, distributing the usage of memory of the fields and parallelizing the workload among the different processors.  

\end{itemize}

While Simflowny 1 supported the Balance Law PDE family through the
Cactus toolkit \cite{cactus, cactusWeb}, Simflowny 2 provides support for SAMRAI mesh
management toolkit~\cite{samrai} for spatial-domain problems and the Boost Graph Library~\cite{boost} for graph problems.

Through SAMRAI, Simflowny 2 supports mesh-based discretization schemes
for PDEs, specifically Finite Difference Methods (FDM) and Finite Volume Methods (FVM). It also
supports particle-based mesh-free Lagrangian methods (e.g.: Smooth Particle Hydrodynamics, SPH), as well as spatial agent
models and Cellular Automata. Notice that SAMRAI only allows models with spatial dimensions
$N>1$. 

Simflowny 2 reflects the parallelization capabilities of both SAMRAI and Boost, and therefore any model developed on Simflowny can generate optimized
parallel code for these frameworks. For efficiency reasons, code responsible for communication is only generated
in the parts of the model which are not local: in PDEs this amounts to
communicate only the fluxes and spatial derivatives while keeping source terms local; in ABM such split is reflected through the gather / update
paradigm: gathers are operations involving neighbours, and are
communicated, while updates do not involve neighbours and
are consequently performed locally to each processor. 

Futhermore, the powerful capabilities of Adaptive Mesh Refinement are available in Simflowny by leveraging its excellent implementation in SAMRAI.

Notice that Simflowny's four stage structure allows us to achieve our stated main goal: a complete split of the physics (the model and the problem) from the numerical 
methods (the discretization schemes) and from the parallelization/distribution issues, which are hidden
in the framework (mesh manager or graph manager). Therefore, we can use the same representation of the discrete problem (which involves the three first stages) in different simulation frameworks, which might allow for higher scalability and efficiency.

Regarding user roles, in Simflowny 2 we consider final users, who are interested in using the platform to build and numerically solve a simulation problem (the user C
pictured in fig. \ref{fig:workflow}), and developer users, who are interested in creating 
new physical models (user A) and discretization schemes (user B). The latter should not be 
confused with the developers of Simflowny, who are interested in adding new 
functionality to the platform itself. 

Simflowny 2 includes example databases for typical physical models and 
discretization schemes. However, in complex simulation scenarios the 
development or tailoring of both physical models and discretization schemes 
by the corresponding experts (A, B) becomes necessary. These elements are 
the starting point for the simulation workflow driven by the final user 
(C).

Simflowny 2 is open source, and it is available in the form of compilable
source code and also as a Docker \cite{docker} container. See 
Simflowny's wiki for further details
\footnote{https://bitbucket.org/iac3/simflowny/wiki/Home}.

The paper is organized as follows. The new GUI architecture is described in Section II. Section III is devoted to the generic PDEs, while Section IV focuses on ABM on spatial domain and on graphs. We finish with some final remarks.

\section{New GUI architecture}
\label{sec:newgui}

In Simflowny, problems, models, and discretization schemes are represented as XML files. 
In the former GUI, the edition of these files was done with many ad hoc
non-reusable GUI components, specifically developed for such documents.

As the families of models supported by Simflowny grew, this approach became unsustainable due to the following issues:
\begin{itemize}

\item It did not take advantage of already existing structure embedded in the XML documents, rather creating a new data structure specifically for the GUI. 

\item It suffered of a lack of coherence, since there was no guarantee --when the GUI components were developed manually-- that similar components were designed in a similar way. In fact, it was possible, specially when the work was done by several developers, that the GUI components differed substantially even if they corresponded to similar information. 

\item It was becoming rather inefficient, with significant development costs due to such heterogeneous designs.

\end{itemize}

These drawbacks led us to completely redesign the GUI to convert it into a specialized XML
editor. Each XML document, accompanied by its schema (XSD) provides all the
necessary information to automatically generate a GUI for editing such document on the fly.

Simflowny's web GUI is composed by two elements: a document manager, which is also the GUI's main page, and a document editor. 

\subsection{Document manager}

\begin{figure*}
  \begin{center}
  \includegraphics[width=\linewidth]{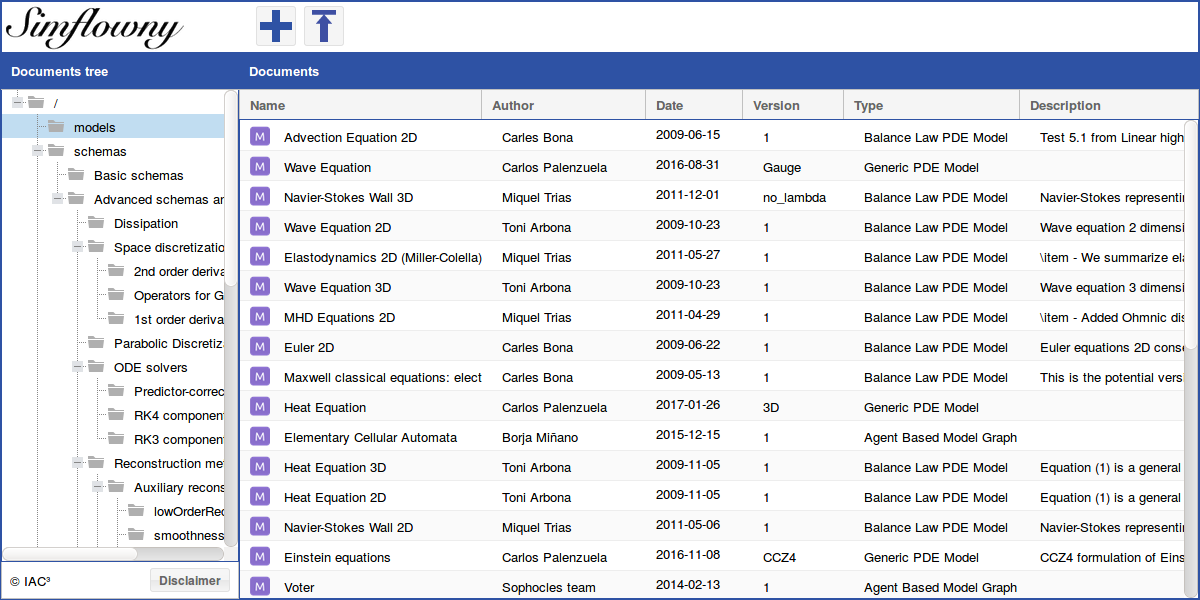}
  \caption{Document manager screen. Notice the tree area on the left. The area on the right shows the available documents in the selected folder.}
  \label{fig:docManager}
  \end{center}
\end{figure*}

The document manager consists of three areas (see figure \ref{fig:docManager}) : i) a toolbar, located at the top, and containing contextual actions for the selected document; ii) a document tree on the left area; and iii) a document list from the selected tree folder on the right area.

 The specific buttons appearing in the toolbar depend on the document actively selected in the document manager. For instance, with no document selected, the only actions are \emph{Create new} and \emph{Import}, as shown in figure \ref{fig:docManager}.

\begin{figure}
  \begin{center}
  \includegraphics[scale=1]{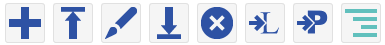}
  \caption{Action bar on the document manager when a problem is selected. The five actions on the left \emph{Create new, Import, Edit, Download, Delete} are common to all documents. The remaining three ones \emph{XML to LaTeX, XML to PDF, Generate Boost Code} are contextual actions, in this case specific to the ABM simulation problem.} 
  \label{fig:barcode}
  \end{center}
\end{figure}

When selecting a document, other buttons appear in the toolbar.
For instance in the case of a document describing an ABM simulation problem, the buttons are those shown in figure \ref{fig:barcode}. The five actions on the left are common to any kind of document; from left to right: \emph{Create new, Import, Edit, Download, Delete}. The remaining three ones are contextual actions, in this case specific for the ABM simulation problem; from left to right: \emph{XML to LaTeX, XML to PDF,} and \emph{Generate Boost Code}. 

The \emph{Create new} button is used to create new documents, whose type is selected on the fly in the button submenu. When clicked, a document editor tab opens with a template document of the selected type. Document edition is explained in detail in the following subsection. The document will be stored in the folder selected in the document tree.
The \emph{Import} button allows uploading documents to the current database. Every document previously exported from Simflowny can be imported.
The \emph{Edit} button opens the document editor with the selected document loaded.
All the documents in Simflowny can be exported. When using the \emph{Export} button the selected documents will be downloaded in a zip, which will contain the selected document and any documents it references. Clicking on the \emph{Delete} button will delete the selected documents.

The \emph{XML to LaTeX} and \emph{XML to PDF} actions generate and download, respectively, \LaTeX and PDF versions of the selected document. The last action generates parallel Boost code for an ABM on a graph (see section \ref{sec:gencode}).

The document tree on the left area is a customizable hierarchical structure. Each element in the tree can be seen as folder where to place the documents. It is possible to reorganize the tree. Right-clicking on any folder opens a contextual menu to \emph{Add} a new folder, or \emph{Rename} or \emph{Delete} an existing folder. It is also possible to move a folder (including subfolders and documents) or documents into another by drag and drop.

The third area in the document manager, at the right of the tree is a document list from the selected tree folder. A double-click on any document will open it in the document editor. The documents can be ordered  and filtered using the header columns.

\subsection{Document editor}

The Document Editor is a specialised XML editor. Underneath,
all documents in Simflowny are XML files, which have a tree
structure. The Document Editor presents such tree structure
in the left panel, which is editable, while a presentation
panel is available on the right, offering a more readable
view of the document being edited. See figures in sections \ref{sec:pde} and \ref{sec:abm}.

\begin{figure*}
  \begin{center}
  \includegraphics[width=\linewidth]{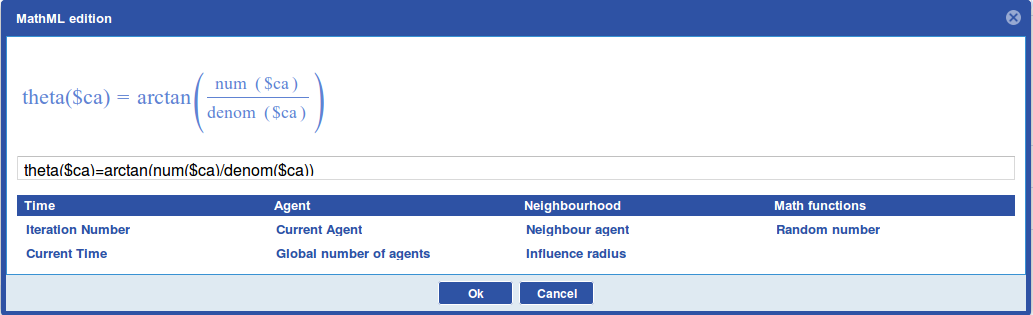}
  \caption{MathML editor. An example of an algebraic expression 
  introduced using the MathML editor. This editor hints the user when 
  writing mathematical expressions, which are entered in the AsciiMath 
  standard in Simflowny. Some SimML instructions can be used in the 
  expressions. The buttons below the editor show all the possible SimML 
  tags available and provide a contextual help for them.}
  \label{fig:mathMLEditor}
  \end{center}
\end{figure*}

The functionalities available for editing documents include:

\begin{itemize}
\item Those addressed to manage the XML elements, among them:
	\begin{itemize}
	\item Decorated versions of the XML tags, to improve
	readability. For instance, the XML tag \emph{addPartialDerivatives} is shown as \emph{Add Partial Derivatives}.
	\item Expanding and collapsing nodes.
	\item Preemptive creation of compulsory elements.
	\item Adding and removing children elements to a parent node. 
	When adding children only
	those allowed by the XSD are shown in contextual menus.
    \item Reordering (moving upwards or downwards) certain elements in a list.
	\item Copy/paste capabilities for SimML elements.
	\end{itemize}
\item Those addressed to manage the content of each XML element, among them: 
	\begin{itemize}
	\item Typing the text content. To enable the edition of an element value, the empty input area next to its label must be clicked.
	\item Support for unicode characters, for instance Greek letters. A graphical tool is provided to facilitate the introduction of such characters. The user may copy a symbol from the pop-up and use it in the editor.
	\item Dropdown lists when the elements can only be chosen from
	a closed list, either defined in the XSD or corresponding to previously defined content in the same XML document.
	\item A specialised editor for SimML instructions.
	\item A specialised math editor for MathML expressions using the AsciiMath standard (see figure \ref{fig:mathMLEditor} for an example)
	\footnote{http://asciimath.org}.
	\item Document cross-reference (hyperlink) graphical tool, used for instance to
	reference a specific model from a problem.
	\item Highlighting missing content to avoid errors.
	\end{itemize}
\end{itemize}

The document is auto-saved when changes are detected and when the browser tab is closed.
At the bottom of the document there is a  button to validate the structure and, as far as possible, the content of the document.
In the same toolbar  area there are two buttons to Expand or Collapse all the nodes in the tree.

\section{Models based on PDEs}
\label{sec:pde}

 In this section it is discussed in detail one of the new families incorporated into Simflowny 2, namely a generic PDE with finite difference discretization. In order to stress the differences with respect to the previous version, we first briefly summarize the previous model for PDEs in Balance Law form. Next we describe the generic PDE family and present an example with the wave equation.
 
 Notice that both families are restricted to  systems with first order derivatives in time, a requirement that can always be satisfied by introducing new evolution fields. This condition allow us to use the Method of Lines (MoL) to ensure stability of the system when converting continuum equations into discrete ones.

\subsection{Pre-existent functionality: PDEs in Balance Law form}

The pre-existent family consists of balance law PDE, namely
\begin{equation}
   \partial_t {\bf u} + \partial_i {\bf F}^i({\bf u}) = 
   {\bf S} ({\bf u}) + 
   \partial_i \left({\bf Q}^{i}_{j}({\bf u})\mbox{ }\partial_k {\bf P}^{kj}({\bf u}) \right)
\end{equation}
where $\bf{u}$ is an array with all the evolved fields, 
${\bf F}$ are the fluxes, ${\bf S}$ the sources, and ${\bf P}$ and ${\bf Q}$ are the algebraic expressions conforming the parabolic terms.
 Since  fluxes and  sources depend only algebraically on the fields, the evolution system is restricted to be first order both in time and space, except for the parabolic terms.
 The advantage of this model is that 
both Finite Difference and Finite Volume methods can
be applied on this form of the equations.

\subsection{New functionality: Generic PDEs}

As it was explained previously, before the generation of the numerical code we must define the model, the problem
and the discretization schemes. This family can explicitly solve any PDE written as
\begin{equation}
   \partial_t \bf{u} = \cal{L (\bf{u},\partial_i \bf{u})}
\label{genericPDE}   
\end{equation}
where $\bf{u}$ is an array with all the evolved fields
and $\cal{L (\bf{u},\partial_i \bf{u})}$ is any operator depending on the fields and its spatial derivatives (of any order). Therefore, the only restriction here is that the system can only involve first time derivatives of the fields.

Arbitrary operators can be constructed by using $n$ recursive rules, which can be written formally as 
\begin{eqnarray}
   {\cal L}_i^{(0)} &=& f_i(\bf{u}) 
\label{recursive_step0}      
\\
                 &...&   
\\   
   {\cal L}_m^{(n)} &=& \sum_{r=0}^{n-1} \sum_i g_i({\bf u}) 
   \prod_j
   \left( \sum_k \partial_k 
    \left( \sum_l {\cal L}_l^{(r)} C^{ijkl}_{m} \right)
    \right)
\label{recursive_stepn}          
\end{eqnarray}

where $\{f_i,..,g_i\}$ are arbitrary functions depending on
$\bf{u}$ but not on its derivatives, and $C^{ijkl}_{m}$
is a generic matrix which in practice will have only
one non-trivial component. Notice that this new family can also be expressed in a logical abstract language as
\begin{eqnarray}
   \partial_t \bf{u} &=& \sum(f(\bf{u}) {\rm D}) 
   \\
   {\rm D} &=& \prod\partial_i(\rm g(\bf{u}) {\rm D}) 
\end{eqnarray}
where, similarly to the former definition, $\{f, g\}$ are arbitrary algebraic functions on $\bf{u}$, and ${\rm D}$ is recursive term allowing for 
a complex set of expressions using differential calculation. Both the algebraic and the recursive terms are optional at every level of the recursion.

Once the model is well defined, we need to set up the problem, namely, the domain of the simulation, the initial and the boundary conditions for the evolved fields, the finalization condition, and the analysis quantities to be computed. 

Before the code generation we must define the discretization
rules. As we mentioned before, we will take advantage of
the theoretical background developed for PDEs by
using the Method of Lines. We have implemented,
as a basic set available in the database provided with Simflowny, some discretization operators  which ensure stability of the discrete
problem (i.e., a third-order Runge-Kutta for the time integration and fourth order centered space discretization), although any other scheme can be defined. By default, it is  assumed a recursive rule such that the $(n)$-order discrete derivative of a field is obtained by applying the discretization operator to the $(n-1)$ derivative of that field.
This procedure provides a straightforward way to compute space derivatives of any order.
Notice however that all these schemes --for the time and space discretization-- can be 
modified freely by setting a different discretization policy. 

\subsection{Generic PDE example : the wave equation}

The use of Simflowny for Generic PDEs is illustrated using the widely known wave equation. The wave equation for a scalar field $\phi$ in 2D
can be written as a system of equations with only first order time derivative, namely
\begin{eqnarray}
   \partial_t \phi &=& K  
\label{wave_eq_phi}   
\\   
   \partial_t K &=& \partial_{xx} \phi + \partial_{yy} \phi 
\label{wave_eq_K}      
\end{eqnarray}

The recursive rules are straightforward here, with
\begin{eqnarray}
   {\cal L}_i^{(0)} &=& \{ \phi, K \} 
\label{recursive_wave_step0}      
\\   
   {\cal L}_j^{(1)} &=& \{ \partial_x \phi , \partial_y \phi\}
                     =  \{ \partial_x {\cal L}_0^{(0)}, \partial_y {\cal L}_0^{(0)}\}
\label{recursive_wave_step1}      
\\   
   {\cal L}_k^{(2)} &=& \{ \partial_{xx} \phi , \partial_{yy} \phi \} = \{ \partial_{x} {\cal L}_0^{(1)},
   \partial_{y} {\cal L}_1^{(1)} \}
\label{recursive_wave_step2}         
\end{eqnarray}

To construct the problem we need to set the domain, the initial data and the boundary conditions.
For this simple example we will set a square domain $[-0.5,0.5]^2$ with an initial parametrized Gaussian profile for the scalar field and periodic boundary conditions.

We will use third order Runge-Kutta for the time integration together with fourth order accurate centered difference operators satisfying the Summation By Parts rule), which ensures stability and convergence of the discrete problem.
The second order space derivatives are calculated by using a 5 point stencil.

\subsection{Model creation}

\begin{figure*}
  \begin{center}
  \includegraphics[width=\linewidth]{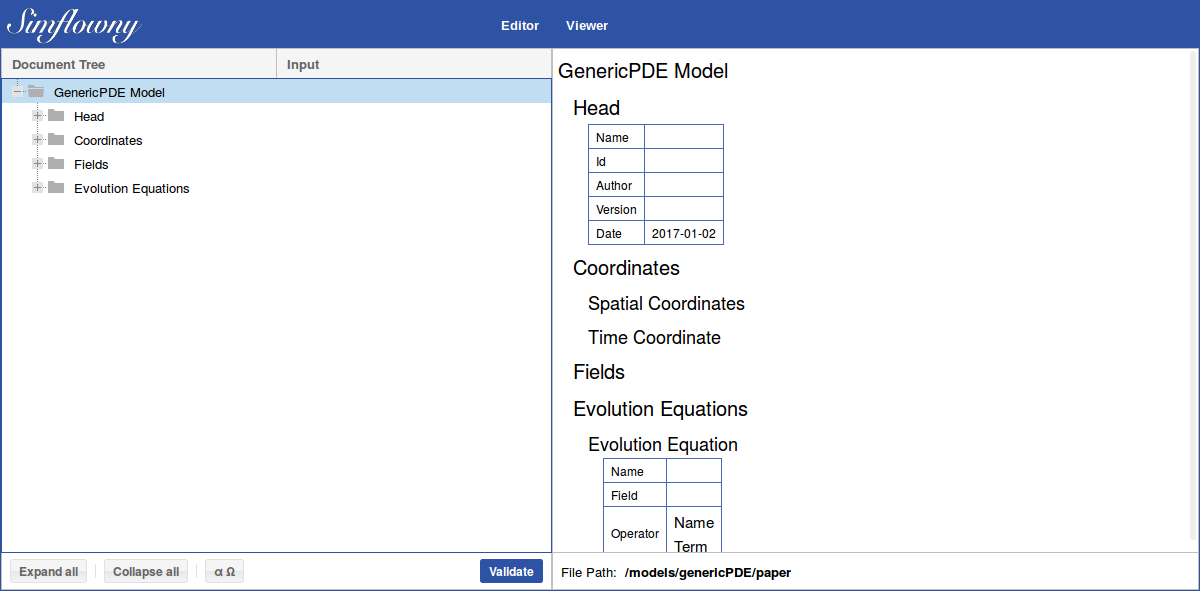}
  \caption{Editor showing the basic skeleton for a generic PDE model}
  \label{fig:editorVacioGenericPDE}
  \end{center}
\end{figure*}

 In the document manager, the user should
select a folder, or create a new one, to store the
model. Next, the user clicks the \emph{plus} button,
expanding a new menu where the option \emph{Generic PDE Model} should be selected. The document editor will open the basic skeleton to define a generic PDE model, as shown in figure \ref{fig:editorVacioGenericPDE}. We will now describe the
different parts of this skeleton.

The head element contains the general description of the model,
which might be filled by the user. By expanding this head, its children are shown (i.e., name, id, author, version, and date).

The first step is to set the spatial dimensions and the names of all the coordinates (i.e., space and time) in \emph{Spatial Coordinates}. By default there is only one \emph{Spatial Coordinate} element. New spatial coordinates can be added 
by the user from the \emph{Spatial Coordinates} element through
its contextual menu. In our particular example, there are two spatial coordinates $\{x,y\}$  and a time coordinate $t$. 

The second step consists in defining the \emph{fields}, that is,  the variables which are going to be evolved in
Simflowny. One \emph{field} is the minimum required to define a model. For the wave equation, the user must define two fields $\{\phi ,K\}$. 

\begin{figure*}
  \begin{center}
  \includegraphics[width=\linewidth]{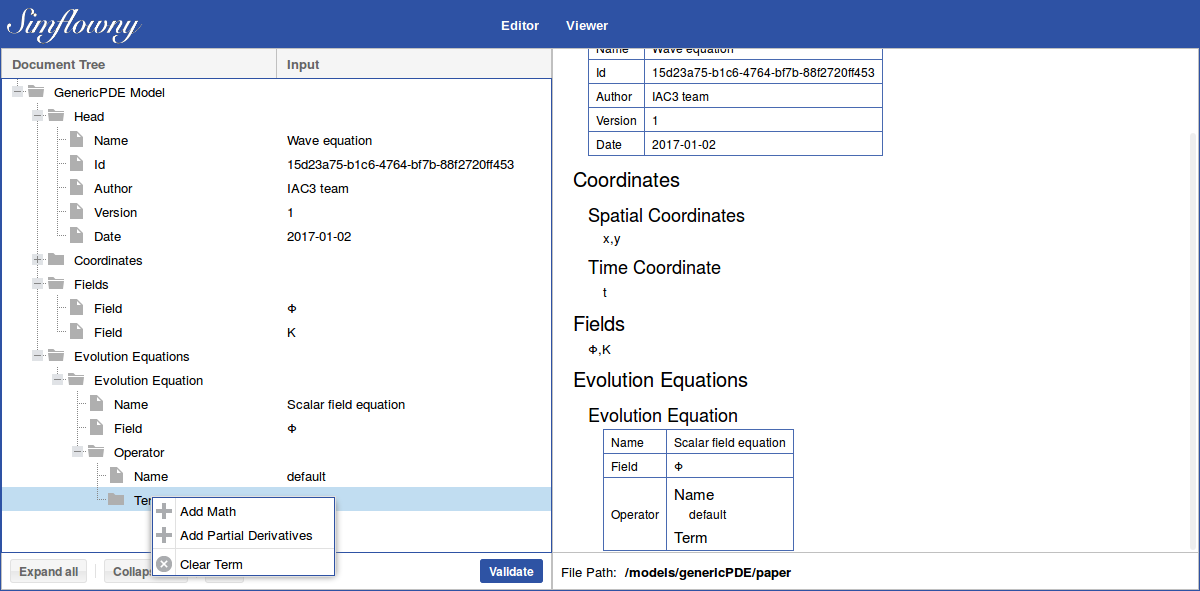}
  \caption{Construction of the wave equation model. In the left panel the contextual menu to add mathematical term for the scalar field 
  equation term is shown.}
  \label{fig:genericPDECreation}
  \end{center}
\end{figure*}

The core of a PDE model relies on the last mandatory section, where the evolution equation (i.e., the operators) associated to every field must be set with a descriptive name.
The generic PDEs have a structure, described by eqs.
\ref{recursive_step0}-\ref{recursive_stepn}, in which the last (n)-recursive step consists on a summation of terms --defined as \emph{Term} elements--, each one with many combinations of derivatives. The terms which are going to be discretized in the same way are grouped by elements called \emph{Operator}.
It is possible to use different discretization schemes for each \emph{Operator}.  As it is shown in figure \ref{fig:genericPDECreation}, the user can introduce both algebraic expressions (\emph{Math}) or derivative ones (\emph{Partial Derivatives}) in the contextual menu of \emph{Term}.

\begin{figure*}
  \begin{center}
  \includegraphics[width=\columnwidth]{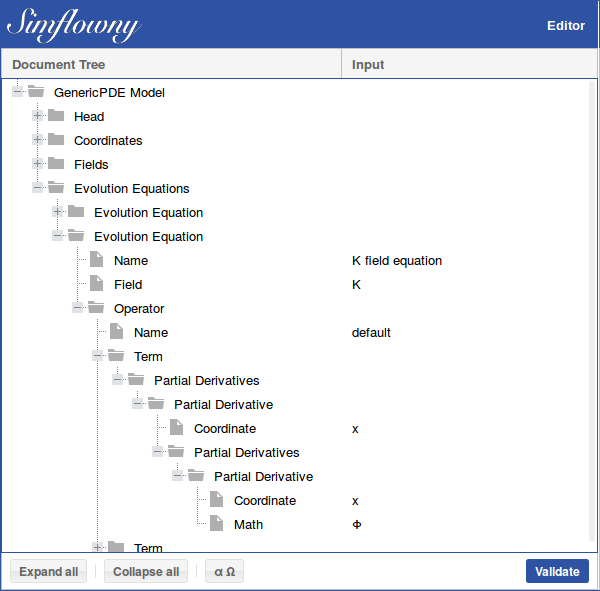}
  \caption{Construction of the wave equation model. Second derivative term introduced.}
  \label{fig:secondDerivative}
  \end{center}
\end{figure*}

\begin{figure*}
  \begin{center}
  \includegraphics[width=\columnwidth]{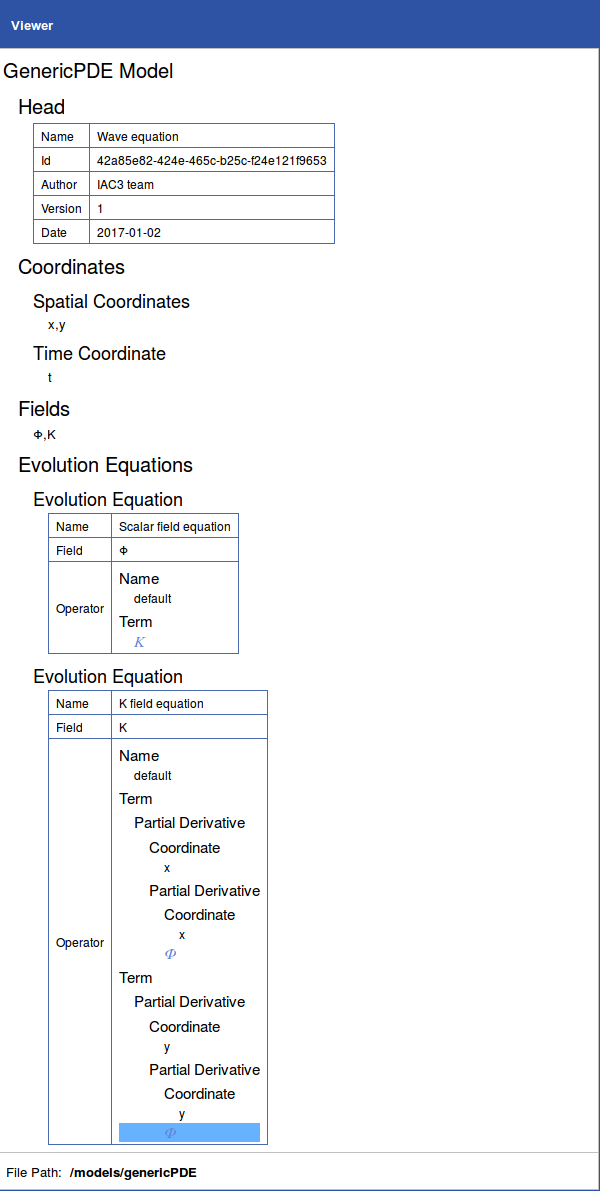}
  \caption{Detail of the right frame of the editor showing the completed wave equation model.}
  \label{fig:genericPDEModelComplete}
  \end{center}
\end{figure*}

In the example with the wave equation, the simplest choice is to use only one operator for both equations, that it will be referred as \emph{default} from now on. For the equation ~\ref{wave_eq_phi} (i.e., the time derivative of the scalar field $\phi$), the \emph{default} operator has only one algebraic expression, \emph{$K$}, to be added as a \emph{Math} element. The mathematical expressions are introduced through the editor explained in section~\ref{sec:newgui}.
The time evolution of $K$, given by eq.~\ref{wave_eq_K}, contains two terms $\{ \partial_{xx} \phi ,\partial_{yy} \phi \}$. 
These terms, which must be introduced as \emph{Partial Derivatives} elements, are calculated by using the recursive rules given in eq.~\ref{recursive_wave_step0}-\ref{recursive_wave_step2} from the top to the bottom level.
For the first term $\partial_{xx} \phi$, the user defines first the most external derivative ${\cal L}_0^{(2)} \equiv \partial_{x} {\cal L}_0^{(1)}$ and sets the coordinate \emph{$x$} in \emph{Partial Derivative}. At the second iteration the next level ${\cal L}_0^{(1)} \equiv \partial_x {\cal L}_0^{(0)}$ is defined following a similar procedure. Finally, at the bottom level the user adds a
\emph{Math} element to define the algebraic function ${\cal L}_0^{(0)} \equiv \phi$. This procedure is displayed in Figure \ref{fig:secondDerivative}.  The process to add the second term \emph{$\partial_{yy} \phi $} is similar to the one explained above. Figure \ref{fig:genericPDEModelComplete} shows the model, completed at this point.

Once the model is finished one can validate it as seen in section \ref{sec:newgui}.

\subsection{Problem creation}

The next stage is to create a problem containing the model, the domain and the initial and boundary conditions.  The problem can be created from the document manager by adding a new document \emph{Generic PDE Problem}.
After setting the header information, the coordinates and fields must be added.

Two parameters $\{a,b\}$ are added for the Gaussian profile of the initial data by using the contextual menu.

\begin{figure*}
  \begin{center}
  \includegraphics[width=\linewidth]{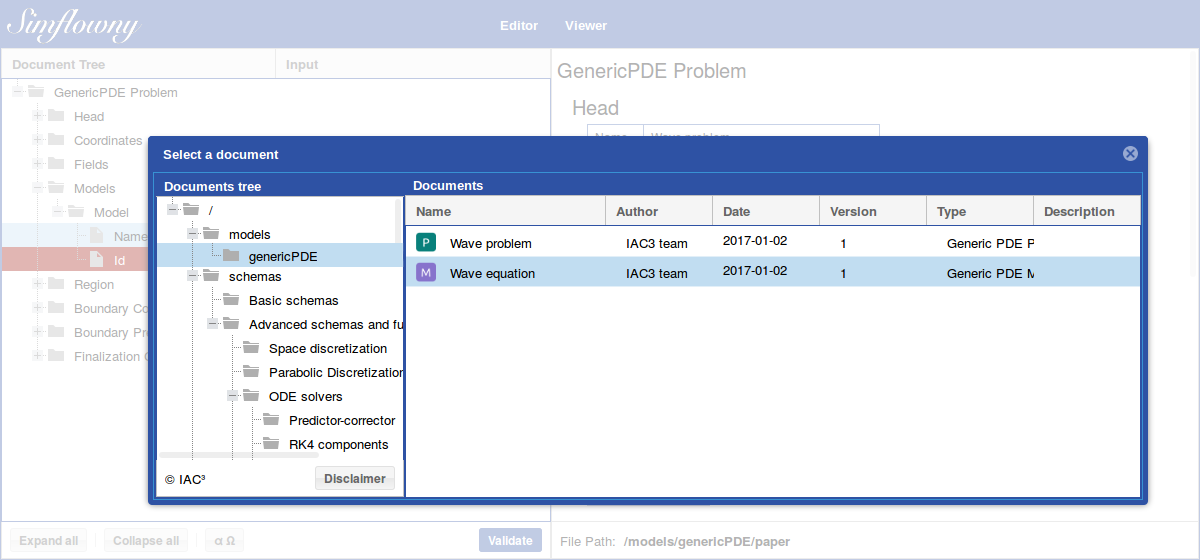}
  \caption{Construction of the wave equation problem. Selection of the wave equation model.}
  \label{fig:modelSelection}
  \end{center}
\end{figure*}

The user may now select the model. This selection is performed through a small document manager pop-up, as seen in Figure \ref{fig:modelSelection}. 

In general, a given problem might include an arbitrary number of models. This is useful, for instance, if these models are going to be applied to different regions of the simulation domain, since Simflowny provides a multi-region capability. By default, the user must define a region covering the full simulation domain. Optionally, one can also define subregions, each one employing different models. 

In our example, the wave problem has a single region. The user must select the wave model as an \emph{Interior Model} in this region.
The spatial domain is set to $x^i \epsilon [-0.5,0.5]^2$ by
assigning a \emph{Coordinate Min} of \emph{$-0.5$} and \emph{Coordinate Max} of \emph{$0.5$} for coordinate \emph{$x$}. The same process is repeated for coordinate \emph{$y$}.

\begin{figure*}
  \begin{center}
  \includegraphics[width=\linewidth]{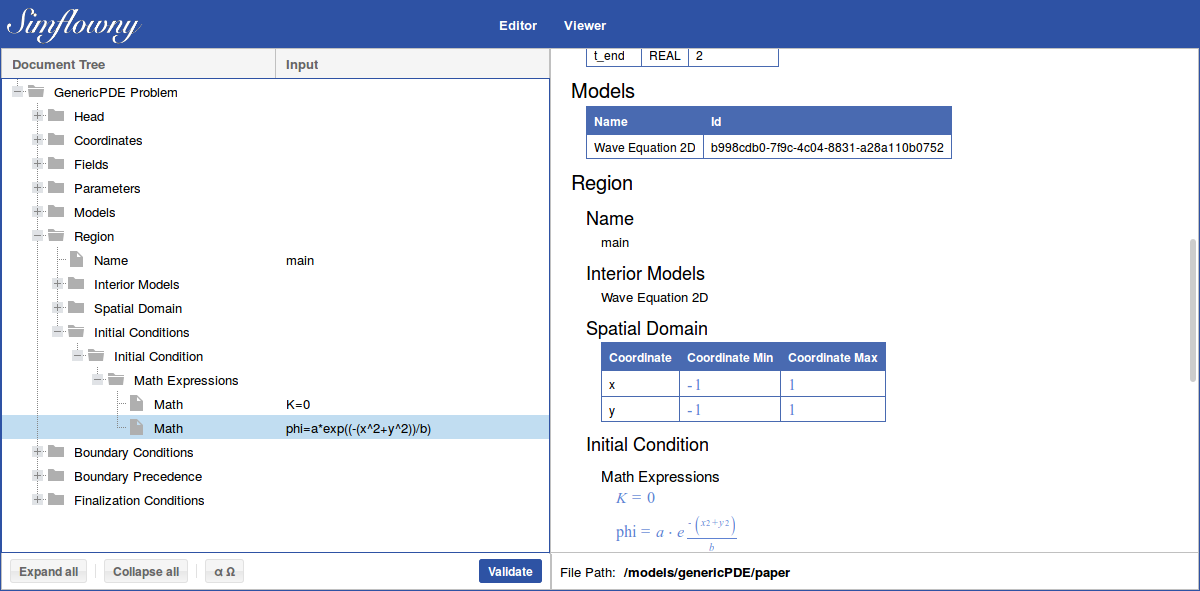}
  \caption{Construction of the wave equation problem. Initial conditions created.}
  \label{fig:initialConditions}
  \end{center}
\end{figure*}

The last compulsory element to set in a region is the \emph{Initial Condition}, which might contain mathematical formulas (\emph{Math Expressions}) and logical conditions (\emph{Apply If}).
In this test case, the initial conditions are:
\begin{equation}
    K=0  ~~~~~,~~~~~ \phi= a \exp(-(x^2+y^2)/b)
\nonumber
\end{equation}
see figure \ref{fig:initialConditions}.

The next step consists in adding the domain \emph{Boundary Conditions}. 
In this element the user sets
first the \emph{Boundary Policy}, which is a mapping of the regions, in this case the region \emph{main},
to a specific \emph{Boundary Condition}. 
Periodic boundary conditions are applied to all the edges and fields by setting the \emph{Boundary Condition} \emph{Type} as  \emph{Periodical} and the 
the \emph{side} and \emph{axis} properties to \emph{all} for all fields, meaning that periodic boundary conditions are applied to all coordinates (i.e., $x$ and $y$) and to both sides (i.e., lower and upper). 

The last setting regarding the boundaries is the \emph{Boundary Precedence}, which indicates in what order overlapping boundary conditions are applied. All the boundary conditions must be added to the \emph{Boundary Precedence} list. In this case it does not matter the order as it does not alter the result.

\begin{figure*}
  \begin{center}
  \includegraphics[width=\linewidth]{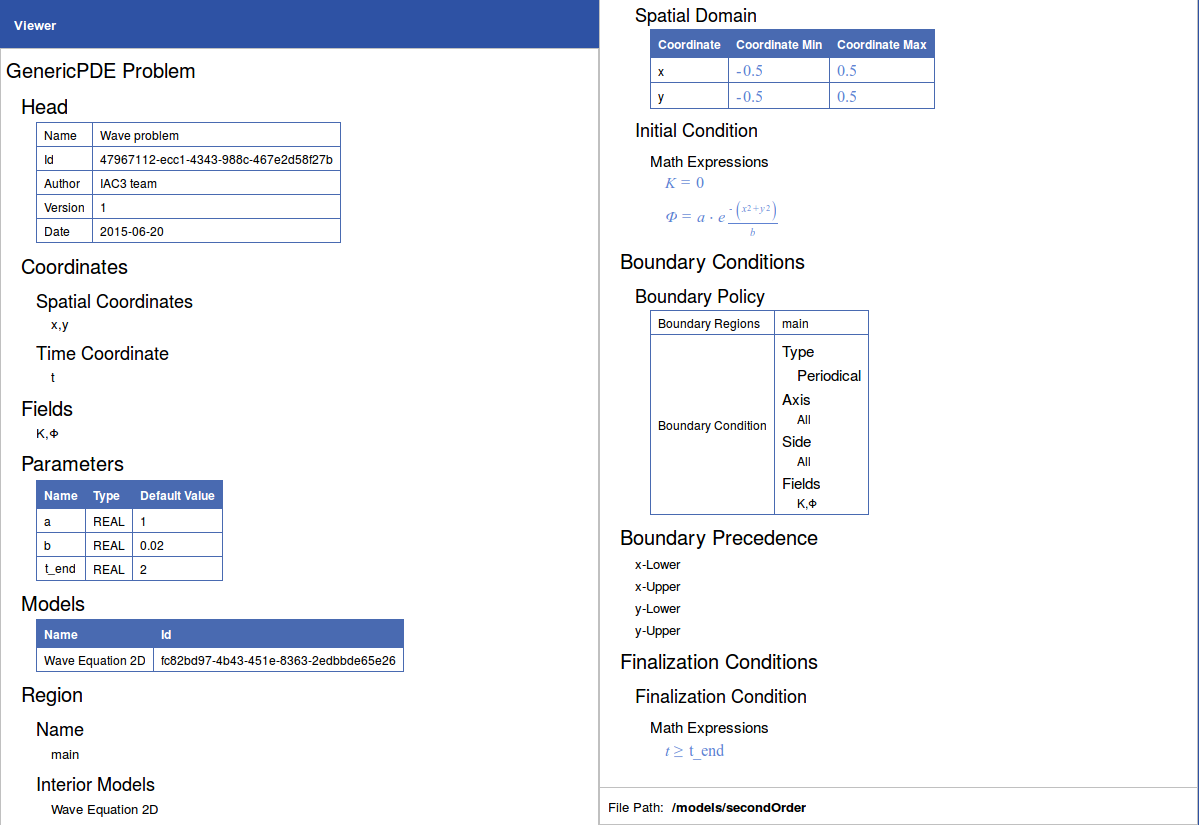}
  \caption{Detail of the right frame of the editor showing the completed wave problem.}
  \label{fig:genericPDEproblem}
  \end{center}
\end{figure*}

The last piece of information needed in a problem is the \emph{Finalization Conditions} element. These conditions have to be logical expressions depending on evolution fields, parameters or variables (i.e., including the coordinates). In this test case, the simulation stops when the evolution time $t$ reaches a certain value provided by the user. Therefore, a new parameter \emph{t\_end} is to be added. Then, the finalization condition \emph{Math} is \emph{$t>=t\_end$}. See figure \ref{fig:genericPDEproblem}, where the complete problem is shown.

\subsection{Problem discretization}

The following stage, after the problem definition, is the discretization. Although the users can create their own, Simflowny provides a set of discretization schemas.

Discretization schemas can be applied to multiple problems and vice versa. Additionally, a schema might be parametrized, so it is possible to use the same schema with different parameter choices to generate different discretized problems. As a consequence, it is necessary a mapping between discretization
schemas and a given problem. This mapping is known as \emph{Discretization Policy} in Simflowny and it is common to all the PDE families. 
Although this policy can be created from scratch, it is easier to generate it from an existing \emph{Generic PDE Problem}: by selecting the problem in the document manager the toolbar shows a button to generate the \emph{Generic PDE Discretization Policy}. 

\begin{figure*}
  \begin{center}
  \includegraphics[width=\linewidth]{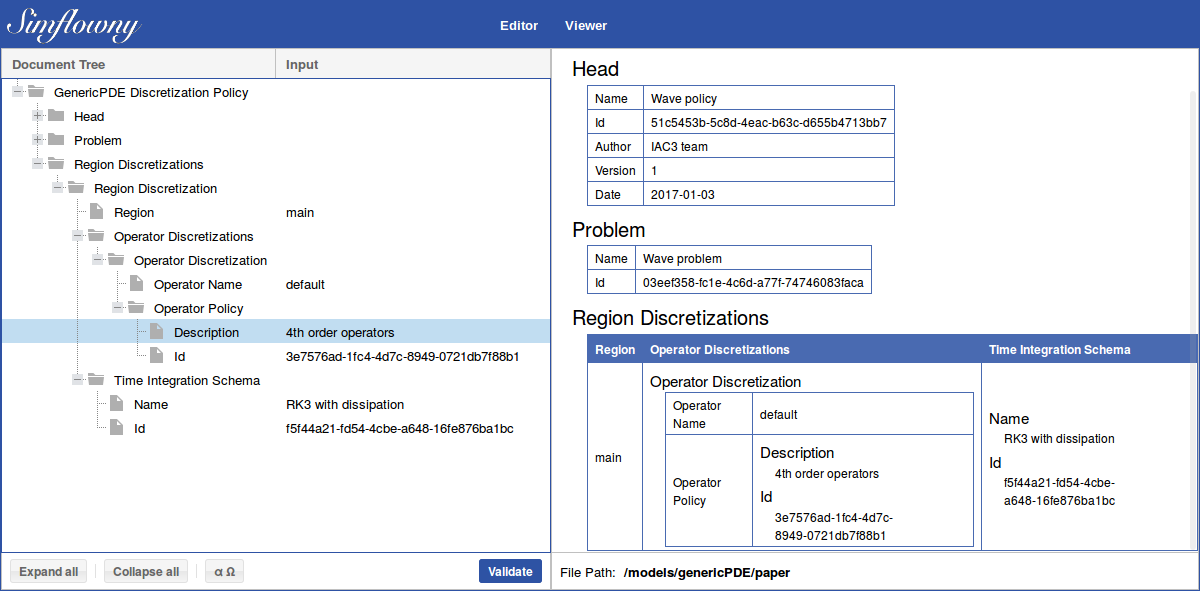}
  \caption{Discretization policy completed.}
  \label{fig:genericPDEDiscretizationPolicy}
  \end{center}
\end{figure*}

The new document automatically incorporates the problem information. Therefore, the only required actions are selecting the space and time discretization schema (i.e., \emph{Operator Discretization} and \emph{Time Integration Schema} respectively) for every region in the problem. In the wave model example, the spatial \emph{Operator Policy} applied in the only region is the \emph{4th Order Operators} schema. The time integration chosen is \emph{RK3 with dissipation}, also available in the database. At this point, the policy is completed,
as seen in Fig.~\ref{fig:genericPDEDiscretizationPolicy}.

Once the discretization policy is defined the user must explicitly discretize the problem. This step is performed in the document manager, after selecting the newly created policy, with the toolbar operation \emph{Discretize Problem}. 

\subsubsection{Generating code}

By using the toolbar, the \emph{Discretized Problem} abstract formulation can be translated into an specific code to be run on a certain simulation platform.
Then, Simflowny starts the code generation process, which finishes with the creation of the \texttt{Generated Code}. The code can be
downloaded in a zip file format, see figure \ref{fig:barcode} and section \ref{sec:newgui}. 

\subsubsection{Running a simulation}

The code compressed in
the zip file can be compiled and executed in any machine with an installed version of SAMRAI.
The unzipped folder will be the
location for the compilation and simulation and contains the source code files, a sample parameter file and a Makefile. 

\begin{figure}
  \begin{center}
  \includegraphics[width=\columnwidth]{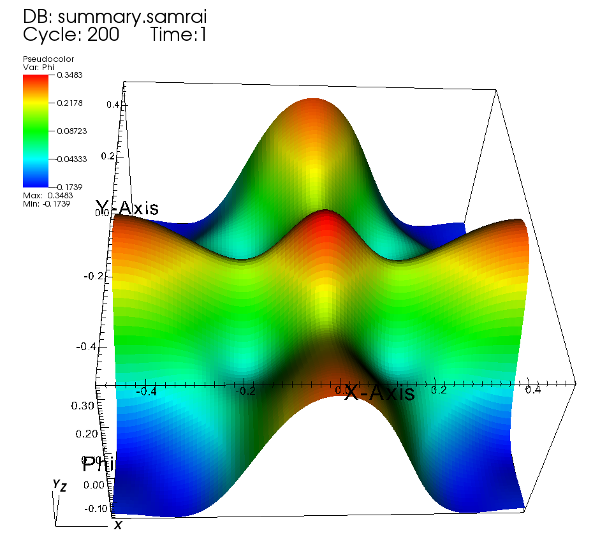}
  \caption{Scalar field results from wave problem simulation.}
  \label{fig:WaveEquationResults}
  \end{center}
\end{figure}

The wave problem example consists on running a simulation up to $t=1$ using a mesh of $100^2$ cells. This can be achieved by modifying the following values in the \texttt{problem.input} file:
\begin{verbatim}
  tend = 1
  ...
  dt = 0.005
\end{verbatim}

The user can compile and launch the simulation by running in a terminal the following commands:

\begin{verbatim}
  make
  ./Waveproblem problem.input
\end{verbatim}

The results are saved by default in the \emph{outputDir} directory and can be visualized with Visit~\cite{HPV:VisIt}. The figure~\ref{fig:WaveEquationResults} shows a snapshot of the results for the scalar field at $t=1$.

\section{New functionality: Agent Based Models}
\label{sec:abm}

This section is comprised of two subsections. First we introduce
Agent Based Models on a graph, and then ABM on a spatial domain.

The difference between ABM on a spatial domain and ABM on a graph
is only the topology of the domain: in the former case agents
have spatial positions and neighbours
of an agent are determined by a distance, while in the latter
agents are the vertices of a graph, and neighbours are determined
by the graph topology. Within the ABM on a spatial domain we
distinguish two subcategories, related to agent motion:
fixed agents (i.e. Cellular Automata) or moving agents.

The ABM models allowed in Simflowny are of the form:

\begin{equation}
\partial_t u =  \mathcal{F}(u) 
\end{equation}

where $\cal{F}$ is a discrete arbitrary algorithm using the properties
of an agent, the properties of its neighbours, parameters, and
coordinates or, in the case of ABM on a graph, characteristics of the graph local topology.
Such algorithm is structured around two types of rules:
\begin{itemize}
\item Collection of neighbour information (gather).
\item Local information processing and updating the local state (update).
\end{itemize}

\subsection{Agent Based Models on a graph}

In this section we validate the new capabilities in Simflowny 2
by building an ABM on a graph model and running a simulation with it. 
We will implement the voter model, a simple model in which
essentially the binary value of an agent is changed according to
the consensus of its neighbours together with a random noise. 

\subsubsection{Model creation}

\begin{figure*}
  \begin{center}
  \includegraphics[width=\linewidth]{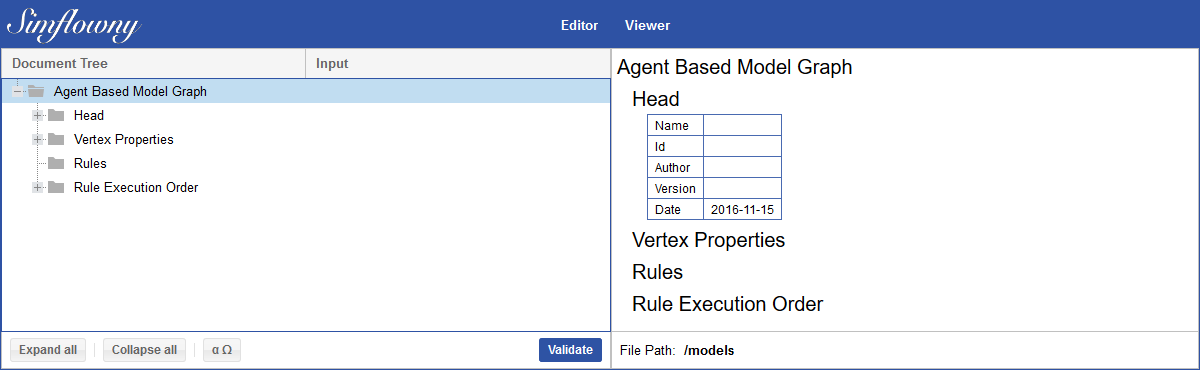}
  \caption{Editor showing the basic skeleton for a model}
  \label{fig:editorVacio}
  \end{center}
\end{figure*}

 In the document manager, the user should
select any folder, or create a new one in order to store the
model there. Next, the user clicks the \texttt{plus} button.
This expands a new menu, where the option \emph{Agent Based Model on graph} should be selected.
Then the document editor will open the basic skeleton for a model
on a graph, as shown in figure \ref{fig:editorVacio}.

Now, the user fills in the general description of the model, contained in
the head tag. By expanding the head, its children tags are shown, namely
name, id, author, version, and date.

The next step consists in defining the \emph{Vertex Properties}. In
Simflowny a \emph{Vertex Property} is a variable which has a specific
value for each vertex in a graph. One \emph{Vertex Property} is the
minimum necessary to define a model.

To build the voter model, the user defines two  \emph{Vertex Properties} called \emph{state} and \emph{acc}. 
\emph{state} stands for the voter model state, which can take two values.  
\emph{acc} is just an auxiliary variable to
accumulate neighbour values.

With this, the user is ready to proceed with the definition of $\mathcal{F}$, which, as stated before, is composed of the rules that tell the model how to change its \emph{Vertex Properties} from
step \emph{n} to step \emph{n+1}. In Simflowny this is specified under
the element \emph{rules}, whose contextual menu allows to
add two children: \emph{Gather Rules} and \emph{Update Rules}.

\emph{Gather Rules} are potentially non-local update operations on a
\emph{Vertex Property} which, for a certain vertex, may involve
information about its neighbours. On the other hand, \emph{Update Rules}
are local update operations, using only information from the same
vertex. While the user can write any algorithm by using only gather rules,
parallel performance and optimization are improved by carefully
distilling local operation rules from operations which may involve
neighbours. This is because no communication between different processors is needed for
update rules, and, as an added bonus, the model description becomes much
clearer when update rules are used wherever possible.

When creating a \emph{Gather Rules} element, one child comes
automatically with it: a \emph{Gather Rule} operation. A model may contain as many as needed, one being the minimum.

A \emph{Gather Rule} operation is defined by \emph{name}, \emph{Vertex Property}, and
\emph{algorithm}, which contains the actual update rule. In the voter model, this first
\emph{Gather Rule}, called here \emph{Acc gather 1}, corresponds to the \emph{Vertex Property} \emph{acc}.

\begin{figure*}
  \begin{center}
  \includegraphics[width=\linewidth]{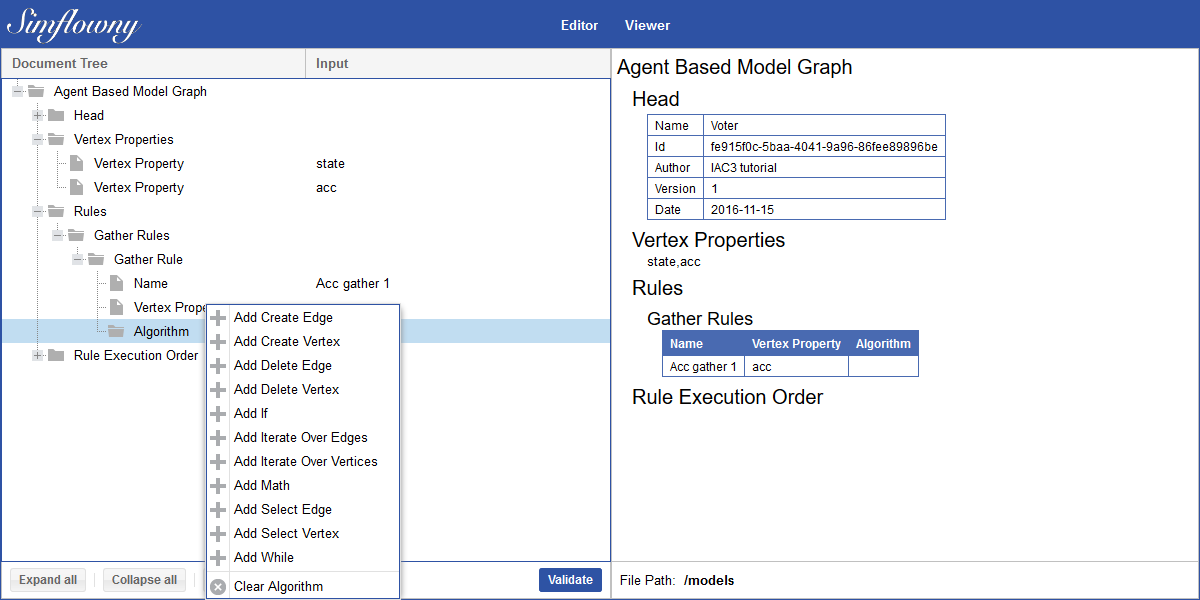}
  \caption{Construction of the voter model. In the left panel the contextual menu to add SimML elements is shown.}
  \label{fig:simmlContextual}
  \end{center}
\end{figure*}

The \emph{algorithm} is defined using SimML and MathML.  The \emph{algorithm}
contextual menu allows the user to select the needed SimML element.  See figure \ref{fig:simmlContextual}.

To continue with the building of the voter model, the user
adds an \emph{Iterate Over Edges} element under the \emph{algorithm} element. This is a loop
over all the edges of a certain vertex.

\emph{Iterate Over Edges} has a child called \emph{Direction Att}  
to indicate the edge direction, \emph{in} in our case.  After this, the user adds a \emph{math} child to
\emph{Iterate Over Edges} with the following content:

\begin{quote}
\emph{acc}(\emph{\$cv}) = \emph{acc}(\emph{\$cv}) + \emph{state}(\emph{\$es}(\emph{\$ce}))
\end{quote}

where
  \emph{\$cv} stands for current vertex,
  \emph{\$ce} stands for current edge,
and  \emph{\$es}(\emph{\$ce}) stands for the neighbour vertex at the other end of the edge (edge source). 



All together, the expression updates the \emph{acc}
\emph{Vertex Property} of each vertex by adding the value of the
\emph{state} \emph{Vertex Property} of the vertices participating in the
edge with the current vertex (i.e., its neighbours). 
This is, \emph{acc} accumulates the values
of \emph{state} for all the (incoming) pairs of a certain vertex.

In a similar fashion, the user adds a mathematical expression in an update rule (\emph{State update 1})
for the \emph{Vertex Property} state:

\begin{quote}
\emph{tmp} = \emph{acc}(\emph{\$cv})/\emph{\$lnoe}\_\emph{in}(\emph{\$cv}) - \emph{\$rnd}\_\emph{uniform}
\end{quote}

In this case we have needed two additional functions:
  \emph{\$lnoe\_in} for the local number of incoming edges (for the current
  vertex) and
  \emph{\$rnd\_uniform} to generate a random uniform variable (between 0 and 1).
Therefore, this expression computes the average value of the incoming
neighbour vertices and then subtracts a random value between 0 and 1.
Notice \emph{tmp} is a temporary variable. 

\begin{figure*}
  \begin{center}
  \includegraphics[width=\linewidth]{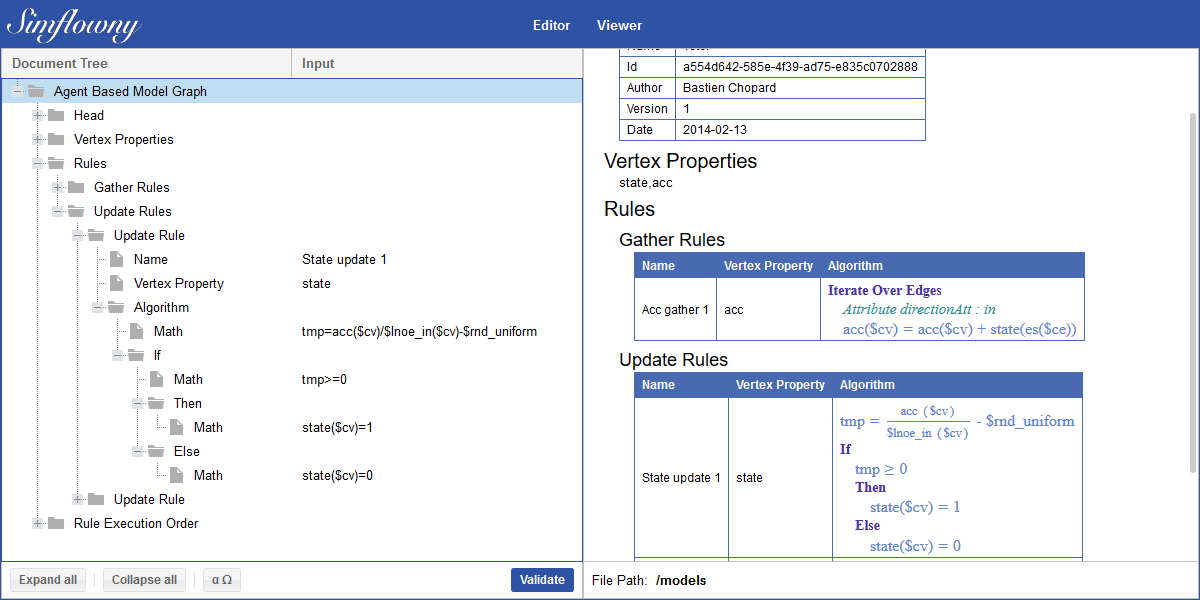}
  \caption{Detail of the \emph{If-Then-Else structure} while building an update rule.}
  \label{fig:updateIf}
  \end{center}
\end{figure*}

At this point, the
value of \emph{state} is not yet updated. To do so, the user should add an \emph{if-then-else}
instruction as a child of \emph{Algorithm}, the \emph{If} condition being  $tmp>=0$, the \emph{Then} statement $state(\$cv)=1$, and the \emph{Else} statement $state(\$cv) = 0$.
Notice that mathematical conditions and statements are introduced as a \emph{math} element.
See figure \ref{fig:updateIf}.

In a similar fashion, the user initializes the \emph{acc}
\emph{Vertex Property} by adding an \emph{Update Rule} called \emph{Acc update 1} and containing the statement $acc(\$cv) = 0$.

Finally, the user just needs to specify
the order in which the different \emph{Gather Rule} and \emph{Update Rule}
instructions must be executed by using the
\emph{Rule Execution Order} element. A \emph{Rule} element, child of \emph{Rule Execution Order},  is added referencing either a \emph{Gather Rule}
or an \emph{Update Rule}. The user chooses the desired element as
the first rule, and then follows on with additional rules
to complete the sequence, in this case: \emph{Acc update 1}, \emph{Acc gather 1}, and \emph{State update 1}

\begin{figure}
  \begin{center}
  \includegraphics[width=\columnwidth]{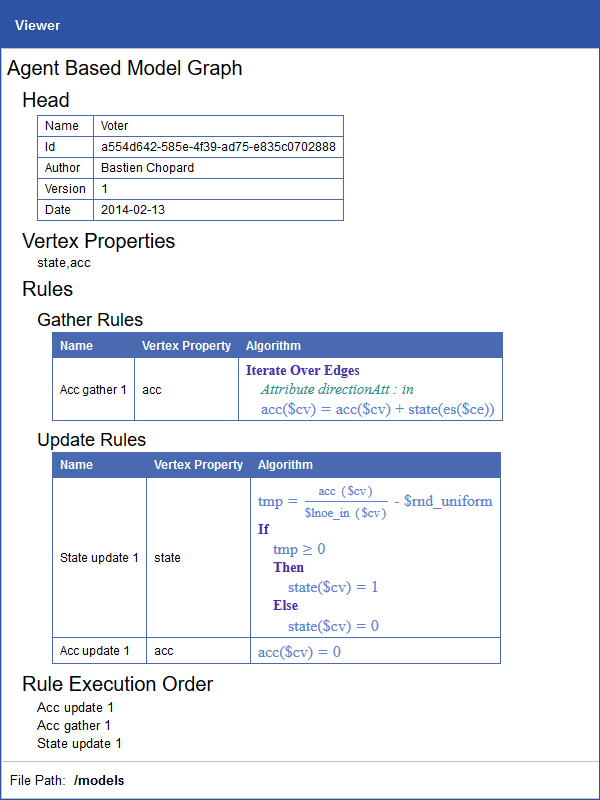}
  \caption{Detail of the right frame of the editor showing the completed voter model.}
  \label{fig:rightframe}
  \end{center}
\end{figure}

Now the definition of the voter model is complete, it can be validated, and it is displayed in the right frame of the editor, as shown in figure \ref{fig:rightframe}.

\subsubsection{Problem creation}

The next stage is the creation of the problem based on the voter model
defined in the previous subsection. 

A problem on a graph domain is
created by selecting the option \emph{Agent Based Problem on graph} from
the add document button.
As in the case of the model, this creates an empty template for the
construction of a problem.

The user fills in the \emph{Head} details and then the
vertex properties: \emph{state} and \emph{acc}, as in the model
(although different names may be used).

\begin{figure*}
  \begin{center}
  \includegraphics[width=\linewidth]{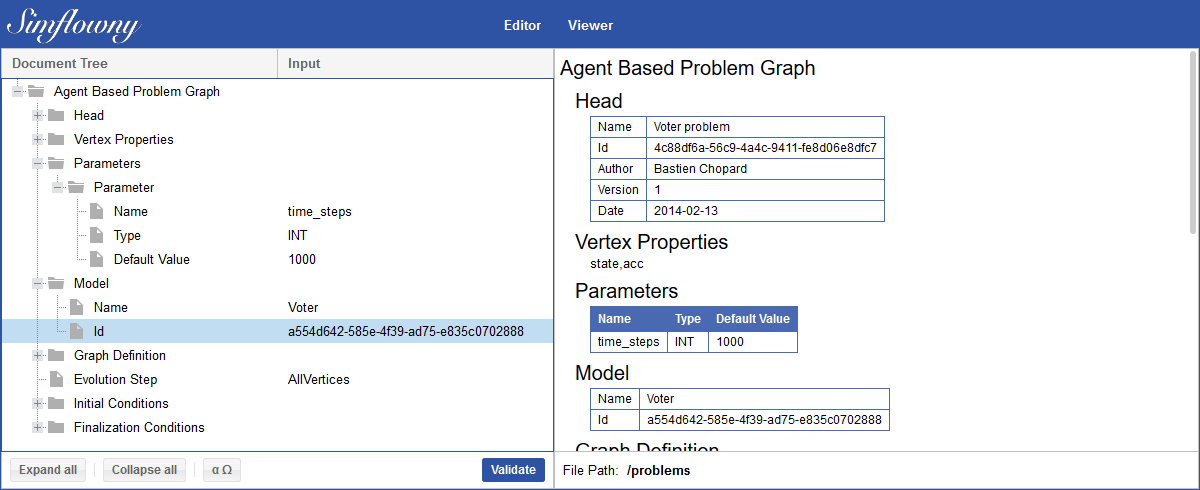}
  \caption{Building the voter problem: head, fields, parameters and model reference.}
  \label{fig:modelid}
  \end{center}
\end{figure*}

As seen in figure \ref{fig:modelid}, the \emph{Parameters} and \emph{Parameter} elements are used to define the \emph{time\_steps} integer parameter.

The following step is referencing a model, using the cross reference graphical tool as explained in section \ref{sec:newgui}. In this example, the user selects the voter model as shown in figure \ref{fig:modelid}.

Now, using the contextual menus provided by the GUI, the user
adds a \emph{time\_steps} parameter of integer type
\emph{INT} and default value \emph{1000}. See figure \ref{fig:modelid}.

\begin{figure}
  \begin{center}
  \includegraphics[width=\columnwidth]{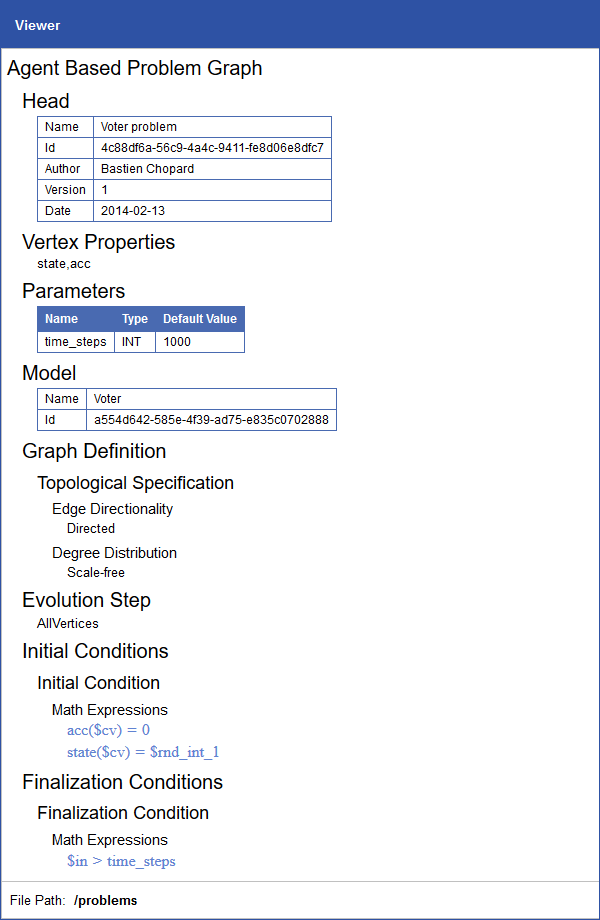}
  \caption{Detail of the right frame of the editor showing the completed voter problem.}
  \label{fig:voterproblem}
  \end{center}
\end{figure}

Simflowny supports either loading a graph from a file or defining it through properties. Such properties allow the user to define the graph as either directed or undirected (\emph{Edge Directionality}) and either random, scale-free or circular (\emph{Degree Distribution}). Our problem is based on a directed random graph. See figure \ref{fig:voterproblem}.

The user can choose to evolve one vertex per time step, or all of them as in this example (\emph{Evolution Step}).  See figure \ref{fig:voterproblem}.

Problems can be optionally endowed with initial conditions, which provide the values for the first step of the simulation. In
this case, a simple mathematical expression suffices:

$state(\$cv) = \$rnd\_int\_1$,

which sets the initial value for state as a random integer
(0 or 1). There is not need to initialize the accumulator, since in the model it is
set to 0 at the beginning of each time step.

Finally, the user needs to establish a finalization condition, that will
specify when to stop the simulation, in this case after a fixed number of time steps.
Notice \emph{\$in}, the iteration number counter, is a
reserved word and its value is automatically increased by one after each
evolution step.
See figure \ref{fig:voterproblem}.

Once finished, the user may click on the button \emph{Validate} to assess the problem integrity.

\subsubsection{Generating code}
\label{sec:gencode}

The button to launch the code generation is located on the action bar on the document manager. 
The user should select the voter problem in the document manager, and click on this button. Then, Simflowny starts the
code generation process. After a few seconds, the code generation
finishes and a new document of type \emph{Generated Code} appears in the document manager. By selecting such document and clicking on the \emph{Download}
button the code is obtained in a zip file, see figure \ref{fig:barcode} and section \ref{sec:newgui}.

\subsubsection{Running a simulation}

The user may download the file to any computer where Simflowny 2 has been properly installed and configured, and unzip it into a folder. This folder will be the
location for the compilation and simulation and contains the source code files, a
sample parameter file and a makefile. 

The use case consists in simulating a random voter model for 10 timesteps, 500 vertices
and 1000 edges. To do so one should set the following values in the \emph{problem.input} file.
\begin{verbatim}
  number_of_vertices = 500;
  vertex_properties=[''state''];
  time_steps = 10;
\end{verbatim}

\begin{figure}
  \begin{center}
  \includegraphics[width=\columnwidth]{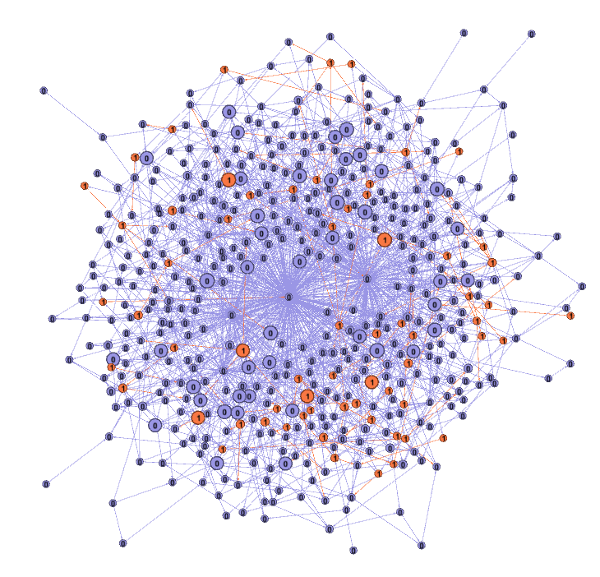}
  \caption{Voter model results.}
  \label{fig:VoterResults}
  \end{center}
\end{figure}

The code compiles using make.
Next, the simulation may be launch  by running:

\verb|./Voterproblem problem.input|

The results are saved by default in \emph{outputDir} directory which can be visualized with any DOT~\cite{dot} viewer. The figure~\ref{fig:VoterResults} shows a snapshot of the results.

\subsection{Agent Based Models on a spatial domain}

In the previous subsection, we have detailed how to create an ABM on a graph,
how to create an associated problem, how to generate its code and how to run it.
Simflowny 2 also includes the new family of ABM on a spatial domain,
which we validate in this section. As many features are common with the
ABM on a graph family, we will directly present an example without
detailing the process of building the model and the problem 

In this example, a two-dimensional flocking model is implemented into Simflowny.
We use the vectorial
noise model of Gregoire
and Chat\'e~\cite{Gregoire2004}. The model is a variation of the
Original Vicsek Model~\cite{Vicsek1995} (OVM),
devised to reproduce the collective motion - or flocking - we
find in many biological and non-biological
systems (see, for instance, references~\cite{Toner1995},\cite{Gregoire2008},
and~\cite{Deutsch2012}and~\cite{Vicsek2012} for reviews).
In these systems long-range orientation order emerges after spontaneous
symmetry breaking.

In the OVM, point-like agents move synchronously in
discrete time steps, with a fixed common speed $v_0$. In 2D the
orientation of agent $\alpha$ is an angle $\theta_{\alpha}$. The evolution
rule provides the new angles at each time step, based on the
angles of the agent's neighbours (agents within a certain
influence radius) in the previous time step. Essentially, the
agent tries to align itself with its neighbours. This alignment is
perturbed by a white noise.

The Gregoire and Chat\'e model is based on the OVM, but
modifies the manner in which noise is incorporated into the model.
They define \emph{vectorial noise} as generated by errors when
estimating interactions, in comparison with the \emph{angular
noise} in the OVM, related to errors in trying to follow
the newly computed direction. Altogether, the update rule for the
Gregoire and Chat\'e model is:

\begin{equation}
\theta^{t+1}_{\alpha} = arg \left[\sum_{\beta \sim \alpha}e^{i \theta^t_{\beta}} + \
\eta n^t_{\alpha} e^{i\xi^t_{\alpha}} \right],
\end{equation}
where $\xi$ is a delta-correlated white noise, $\eta$ represents noise-strength, and $n$ is the number of neighbours. The sum is made over all
the neighbours ($\beta$) of agent $\alpha$.

The solution of this system ranges from nearly complete orientation order
for low noise intensity to random orientation for high noise intensity.
These phases are separated by a novel phase transition. The solutions
near the transition point are characterized, for a wide spectrum of
parameters, by ordered moving structures
(bands) separated by disordered interband regions.

\subsubsection{Generating code}

Assuming we already have the model and problem introduced in Simflowny,
rather similarly as in the voter model case, the code generation can be launched using the button from the action bar appearing when the problem is selected.
The user should select the Collective Motion problem in the document manager, and click on this button. Then, Simflowny starts the
code generation process. After a few seconds, a new document representing the generated code appears in the document manager. It can be downloaded by selecting such document and clicking on the \emph{Download}
button.

\subsubsection{Running a simulation}

After downloading the code, the user may unzip it and run a simulation. The unzipped folder will be the
location for the compilation and simulation and contains the source code files, a
sample parameter file and a makefile. 

The use case consists in simulating a Collective Motion for 10 timesteps in a $[0,100]^2$ domain. To do so one should modify the following values in the \emph{problem.input} file.
\begin{verbatim}
  time_steps = 10
  ...
  dt = 1
  ...
  x_up = 100.0, 100.0
\end{verbatim}

\begin{figure}
  \begin{center}
  \includegraphics[width=\columnwidth]{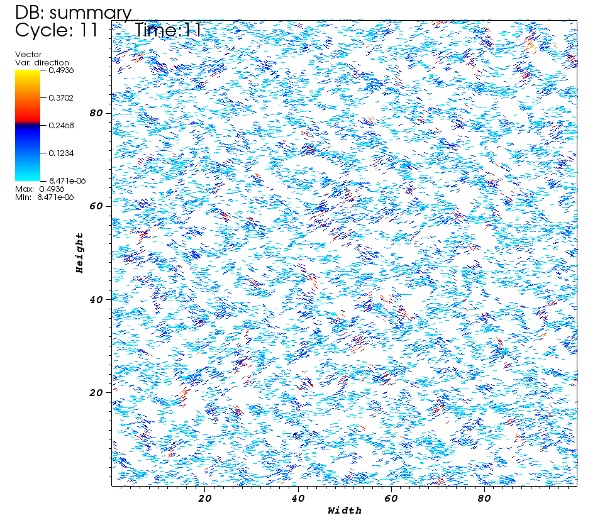}
  \caption{Collective motion results.}
  \label{fig:CollectiveMotionResults}
  \end{center}
\end{figure}

The code compiles using make.
Next, the simulation may be launch  by running:

\verb|./Collectivemotionproblem problem.input|

The results are saved by default in \emph{outputDir} directory and can be visualized with Visit~\cite{HPV:VisIt}. The figure~\ref{fig:CollectiveMotionResults} shows a snapshot of the results.

See \cite{Arbona2014} and \cite{Arbona2015} for the use of Simflowny's flocking model in more complex scenarios.

\section{Conclusions and future work}

We have presented version 2 of Simflowny, a platform for scientific
modelling and simulation on parallel frameworks. Alongside the description of a new GUI architecture to allow for a more agile inclusion of new paradigms of scientific models, we have presented two newly
available families: generic PDE (PDE written in free-style form, in contrast with the
Balance Law form), and ABM, either on a spatial domain or on a graph. We have
illustrated the new families with simple models, but there are a number
of more advanced models available, either to use directly or as a starting point 
to more tailored models. These include Navier-Stokes Equations (Hydrodynamics), 
Maxwell Equations (Electromagnetism), Einstein Equations (General Relativity), 
Brusselator, or Ising models. All of them are included in the basic model 
library of Simflowny. 

In the future, we will continue expanding the database of available models and include new families. We are also working towards supporting unstructured meshes and allowing for
discretization schemes on such meshes (such as Finite Element Methods). 

\section{Acknowledgements}
The research leading to these results
has received funding from the European
Union Seventh Framework Programme (FP7/2007-2013) under grant agreement no 317534 (the Sophocles project).

\bibliography{iac3}

\appendix

\section{SimML sample set}

SimML is Simflowny's Simulation Markup Language, a tag-based specification that allows the creation of complete algorithms to manage PDE and ABM models. Its tags can be mixed with MathML notation to conform the algorithm. 
SimML tags can provide the functionality to
manage the algorithm flow,
use model elements,
change the context of instructions, and
use specific values of the problem.
Some tags can be expressed in infix notation to allow for its inclusion into MathML expressions.

For instance, two typical tags are \emph{if} and \emph{while}. The first introduces a conditional in the algorithm flow based on a logical expression, introduced by the user using MathML. The second creates a loop, which repeats its internal instructions until a logical condition is accomplished.

PDE problems commonly use \emph{boundary} to invoke the boundary conditions (whatever they are), while balance-law PDEs use \emph{flux} and \emph{sources} to call the evolution equation fluxes and sources respectively. They can use \emph{current cell}, \emph{increment coordinate} or \emph{decrement coordinate} to access variables in specific cells.

ABM most common tags  are \emph{current agent} and \emph{current vertex} to acces variables in the agent and vertex, or \emph{edge source} and \emph{edge target} to access variables from vertices connected to the current one.

The  tags \emph{iterate over cells}, \emph{iterate over agents} and \emph{iterate over vertices} are used to sweep over all the cells (PDE), agents (ABM on spatial domain) and vertices (ABM on graphs) respectively and execute the instructions within the tag. To iterate over neighbour agents or vertices, the tags \emph{iterate over interactions} and \emph{iterate over edges} are appropriate.

There are also tags to get values of \emph{field}, \emph{spatial coordinate}, \emph{time coordinate} or \emph{iteration number} among others. Nondeterministic behaviour can be introduced through a random number generator when \emph{random number} tag is set.

Some global accounting information can be obtained using the self-explanatory tags \emph{global number of vertices}, \emph{global number of agents}, \emph{global number of edges} and \emph{local number of edges}.

In ABM on spatial domain, the tag \emph{neighbour agent} is used to access variables of neighbour agents (i.e. those which are in the vicinity of the current agent and which are listed through the \emph{iterate over interactions}  tag).

Some of these tags appear in the examples presented in the paper. See for instance figure \ref{fig:updateIf}.

For a full reference see Simflowny SimML documentation\footnote{https://bitbucket.org/iac3/simflowny/wiki/Simml}.
\end{document}